\documentclass[ba,preprint]{imsart}
\pubyear{0000}
\volume{00}
\issue{0}
\doi{0000}
\arxiv{}
\firstpage{1}
\lastpage{1}

\usepackage{amsthm}
\usepackage{amsmath}
\usepackage{natbib}
\usepackage[colorlinks,citecolor=blue,urlcolor=blue,filecolor=blue,backref=page]{hyperref}
\usepackage{graphicx}

\startlocaldefs
\endlocaldefs

\usepackage{amsfonts}
\usepackage{multicol, booktabs, bbm, color, xcolor}
\usepackage{caption}
\usepackage{subcaption}
\graphicspath{{./img/}}

\begin{document}

\begin{frontmatter}
\title{Incorporating prior information into distributed lag nonlinear models with zero-inflated monotone regression trees
}
\runtitle{Monotone distributed lag nonlinear models}

\begin{aug}
\author{\fnms{Daniel} \snm{Mork}\thanksref{addr1}\ead[label=e1]{dmork@hsph.harvard.edu}}
\and
\author{\fnms{Ander} \snm{Wilson}\thanksref{addr2}\ead[label=e2]{ander.wilson@colostate.edu}}

\runauthor{D. Mork, A. Wilson}

\address[addr1]{Harvard T.H. Chan School of Public Health, 677 Huntington Ave, Boston, MA, USA 02115 
    \printead{e1} 
}

\address[addr2]{Colorado State University, 1877 Campus Delivery, Fort Collins, CO, USA 80523\\
    \printead*{e2}
}


\end{aug}

\begin{abstract}

In environmental health research there is often interest in the effect of an exposure on a health outcome assessed on the same day and several subsequent days or lags. Distributed lag nonlinear models (DLNM) are a well-established statistical framework for estimating an exposure-lag-response function. We propose methods to allow for prior information to be incorporated into DLNMs. First, we impose a monotonicity constraint in the exposure-response at lagged time periods which matches with knowledge on how biological mechanisms respond to increased levels of exposures. Second, we introduce variable selection into the DLNM to identify lagged periods of susceptibility with respect to the outcome of interest. The variable selection approach allows for direct application of informative priors on which lags have nonzero association with the outcome.  We propose a tree-of-trees model that uses two layers of trees: one for splitting the exposure time frame and one for fitting exposure-response functions over different time periods. We introduce a zero-inflated alternative to the tree splitting prior in Bayesian additive regression trees to allow for lag selection and the addition of informative priors. We develop a computational approach for efficient posterior sampling and perform a comprehensive simulation study to compare our method to existing DLNM approaches. We apply our method to estimate time-lagged extreme temperature relationships with mortality during summer or winter in Chicago, IL. 
\end{abstract}

\begin{keyword}[class=MSC]
\end{keyword}

\begin{keyword}
\kwd{Bayesian additive regression trees}
\kwd{monotone}
\kwd{distributed lag}
\kwd{variable selection}
\kwd{informative priors}
\end{keyword}

\end{frontmatter}

\section{Introduction}

In many applications, there is interest in estimating the relationship between a predictor and an outcome when there is substantial prior information about the exposure-response relationship. Including prior information into a statistical analysis can increase the precision of an estimator. In this paper, we consider estimation of the relationship between temperature exposure and mortality due to exposure on the same day and 20 subsequent days. In the environmental epidemiology literature this is commonly referred to as lagged effects. There is substantial prior research that confirms high summer temperatures and low winter temperatures are both associated with increased risk of mortality, the relationship between high or low temperatures, and mortality are each monotonic and exposures with the largest impact on mortality occur on the same day and the previous 3 to 10 days \citep{baccini2008heat,yu2012daily,ragettli2017exploring}. However, there is a lack of appropriate methods to estimate lagged health effects with shape constraints or informative priors on the length of the lagged association. 

When estimating the association between an exposure and an outcome on each of the days following exposure, henceforth lags, the most common statistical approach is a distributed lag model (DLM). DLMs are commonly used in time-series studies to estimate the lagged relationships between an exposure and a health outcome \citep{Schwartz2000a} and in the analysis of perinatal cohort studies to identify time periods during pregnancy when exposures are related to changes in birth or children's health outcomes \citep{Hsu2015}. In a DLM, an outcome $y_t$ at time $t$ is regressed on repeated measurements of past exposures, $x_{t-l}$, for lagged times $l=0,\ldots,L$. Such models assume a linear exposure-response relationship at each lag with lag-specific slope. Exposures measured at fine temporal resolution tend to be highly correlated. To reduce the effect of multicolinearity, DLMs are typically constrained so that the lagged relationships vary smoothly in time. Methods to constrain DLMs include splines \citep{Zanobetti2000}, principal components \citep{Wilson2017a}, Gaussian processes \citep{Warren2012b}, and regression trees \citep{Mork2022EstimatingPairs}. To allow for nonlinear associations between a lagged predictor and an outcome, \cite{Gasparrini2010} proposed distributed lag nonlinear models (DLNMs) that allow for a smooth, nonlinear exposure-lag-response function at each time point using a bi-dimensional spline basis. \cite{Gasparrini2017} later extended DLNM to penalized regression splines that reduce sensitivity of model selection. \cite{Mork2022TreedModels} proposed a regression tree DLNM (TDLNM) approach as an equally powered and more precise alternative. Both DLM and DLNM are now standard tools in the environmental epidemiology literature.

Imposing monotonicity in a distributed lag nonlinear model is appealing because it forces the resulting estimate to correspond with the underlying biological belief that an increase in exposure will not result in an improved health outcome. Yet, there are no existing methods to estimate an exposure-lag-response function subject to a monotonicity constraint. Some previous work has estimated monotone exposure-response functions for exposure to air pollution or weather and a health outcome assessed on the same day \citep{Powell2012, Wilson2014c}, and  there is a rich statistical literature on shape constrained regression for a general regression function. Approaches for shape constrained regression in a general regression setting include piecewise linear functions \citep{Hildreth1954, Brunk1955}, kernel smoothers \cite{Mammen1991}, a large number of spine based approaches \citep{Ramsay1988,Neelon2004,Meyer2008,Wang2008a,Meyer2011, Meyer2012,Powell2012} and  Bernstein polynomial methods \citep{Chang2005,Chang2007,Curtis2011,Wilson2014c,Ding2016,Wilson2020a}. \cite{Chipman2021} proposed a monotone regression model based on the popular nonparametric Bayesian additive regression trees (BART) framework \citep{Chipman2012} that constrains a general regression function to be monotone with respect to some or all predictors. None of these methods are applicable to repeated measures of exposure.

A second appealing area to apply prior information is on which lags have a nonzero exposure effect. Incorporating prior information on lag selection can improve precision in the identified lags during which exposure is associated with an outcome, which are critical to targeting public health interventions. These time periods are of particular interest when estimating the association between maternal exposure to air pollution during pregnancy and birth outcomes where time periods of interest are referred to as critical windows of susceptibility and hypothesized to correspond to sensitive stages of fetal development \citep{Wright2017}. There is limited work on including prior information on which lags represent susceptible periods. To incorporate biological information into the analysis of maternal exposure to air pollution and childhood asthma,  \cite{Hazlehurst2021} used average exposure over clinically defined developmental periods and assessed which periods demonstrated the greatest association between exposure and asthma risk. Using average exposure over predefined windows can cause bias \citep{Wilson2017PotentialHealth} and there are no existing methods to add prior information on which lags have nonzero association with the outcome in the distributed lag framework. For linear DLMs, previous work focuses on adding prior knowledge about the smoothness of lag-to-lag variation in the magnitude of the effect in linear DLMs. This can be achieve through Bayesian priors \citep{Heaton2012} or a priori knot selection in spline models \citep{Gasparrini2016}. In a time series study it is typically to allow more flexibility on the shape of the distributed lag function for short lags during which the majority of the exposure effect is posited to occur \citep{Gasparrini2016}. Yet, these previous methods do not specify a prior on whether there is an effect at specific lags or not, with the exception of smoothly going to zero as the lag increases in some cases \citep{Heaton2012}.

Adding prior information on which lags are associated with the outcome is difficult with most existing models because the probability of a nonzero effect at each lag is not directly parameterized in most distributed-lag-type models. This not only hinders assigning prior probability of a nonzero effect but also reduces interpretability and inference on lag selection. \cite{Warren2020CriticalBirth} proposed a Bayesian variable selection approach to identify time periods where there is a nonzero linear relationship. This approach directly parameterizes inclusion or exclusion of time periods, but does not apply prior information to the inclusion probabilities. For a nonlinear DLNM, time periods are typically identified using confidence or credible intervals to compare expected outcomes compared to a reference outcome, such as zero or median exposure. This both results in a multiple testing issue and is unsatisfying because comparing to a single reference exposure value may miss associations that are only present over certain exposure ranges such as very high exposures. Hence, there is a need for DLNM models that can allow for both assigning prior information to which lags have nonzero effect and for direct inference on time periods when there is a true exposure-response function.

In this paper, we propose a monotone model using  BART-style models that is specifically tailored to repeated measures of exposure using the DLNM framework. Our approach, which we call monotone-TDLNM throughout this paper, utilizes a nested tree BART framework \citep{Chipman2012,Mork2021HeterogeneousPollution}. Specifically, we employ an ensemble of regression trees that subdivide the lagged time periods of exposure. Within a single time-tree, each terminal node corresponds to a mutually exclusive time period of exposure which is affiliated with a nested regression tree that specifies a monotone exposure-response function for the exposure observed in the given time period.  Hence, the exposure-trees are nested in the time-tree to make a tree of trees. To ensure monotonicity we implement a constraint on terminal node parameters of the exposure-trees using a transformation that allows for Gibbs sampling in a hybrid Markov chain Monte Carlo (MCMC) approach to posterior sampling. We introduce variable selection into the DLNM through a zero-inflated alternative to the standard BART tree splitting prior. By combining the zero-inflated splitting prior in an ensemble regression trees setting we are able assign prior probabilities of a nonzero effect at each lag and to make direct inference on the time periods of susceptibility to exposure in our monotone model. The elegance of the nested tree approach is that the exposure-response relationship at each time point is specified by an ensemble of univariate regression trees--nested trees that only splits on exposure concentration. By using univariate trees we can efficiently add monotonicity constraints and selection at each time point.

We demonstrate the advantage of our monotone-TDLNM compared to unconstrained TDLNM and penalized spline DLNMs through a simulation study. Specifically, we show the monotonicity constraint results in more precise estimates of both the exposure-lag-response function and of time periods when there is a nonzero association. We apply monotone-TDLNMs to a reanalysis of temperature exposure and mortality in a time-series study from Chicago, Illinois, USA. Previous analyses of these data have looked broadly at temperature, which tends to have an ``inverted-J'' shape with both high summer temperatures and low winter temperatures being associated with increased mortality. We consider separate analysis of summer and winter temperatures. We separately estimate the monotonic relationship between high summer temperatures and mortality and between low winter temperatures and mortality. Software is made available in the {\tt R} package {\tt dlmtree} (\url{github.com/danielmork/dlmtree}). Code to replicate our simulation and data analysis are available at \url{https://github.com/danielmork/monotone_dlnm}.



\section{Distributed lag nonlinear models}
We begin by introducing the DLNM in the context of a time-series study on temperature related mortality. Let $y_t$ represent the observed mortality count at time $t$. We are interested in how extreme temperature during preceding days is related to changes in $y_t$. Denote $x_t,x_{t-1},\ldots,x_{t-L}$ to be the temperatures during the same day as the outcome and the previous $L$ days. The time-lagged relationship between temperature and mortality is described through the equation
\begin{equation}
    g\left[\mathbb{E}(y_t)\right]=\sum_{l=0}^L w(x_{t-l},l) + h(t;\boldsymbol{\zeta})
\end{equation}
where $g(\cdot)$ is a link function; $h(t;\boldsymbol{\zeta})$ is a function of time parameterized by $\boldsymbol{\zeta}$ and $w(x_{t-l},l)$ is a nonlinear exposure-lag-response function for characterizing the effect of temperature $x_{t-l}$ on mortality $l$ days after exposure. We assume $h(t;\boldsymbol{\zeta})$ contains an intercept and in some cases may contain other covariates such as day of week.

Two assumptions are often imposed when estimating $w$. First, it is assumed that $w$ varies smoothly across the range of $x_{t-l}$ at each lag time $l$. This assumption follows from biological plausibility, that an exposure-response will exhibit a smooth trend across the range of exposure concentration. The second assumption is that $w(\cdot,l)$ is similar for proximal lags.  This assumption is both biologically motivated and statistically practical. From a biological perspective it is assumed that exposure on proximal days will have a similar effect on the outcome. Statistically, there typically exists high autocorrelation between exposure measurements taken close in time and some form of regularization is needed to reduce the effects of multicolinearity in the predictors and ensure biologically plausible estimates of the exposure-lag-response function. Regularization in the lag dimension can take the form of a smoothness constraint or piecewise smooth or constant parameters over a small number of time segments. Combined, these assumptions result in a smoothly or piecewise smoothly varying exposure-lag-response function.

\section{Nested tree framework for DLNM}
\begin{figure}[!ht]
    \centering
    \includegraphics[width=10cm]{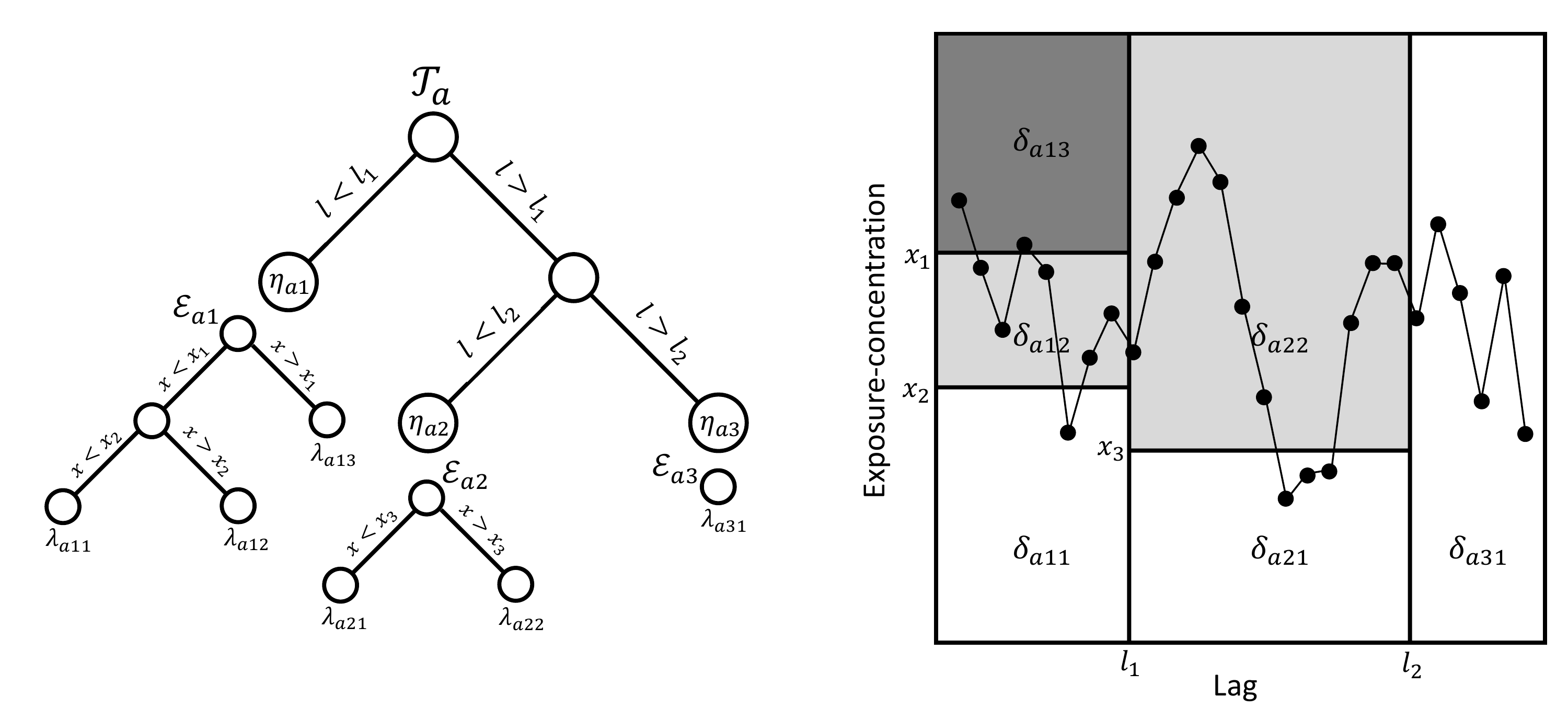}
    \caption{Diagram of the nested tree DLNM framework. On the left is a single binary tree $\mathcal{T}_a$ from the ensemble of trees, $a=1,\ldots,A$. Tree $\mathcal{T}_a$ partitions the lag dimension via terminal nodes $\eta_{ab}$, $b=1,2,3$, that partition the time dimension. The nested trees, $\mathcal{E}_{ab}$, and corresponding terminal nodes, $\lambda_{abc}$, partition the exposure-concentration dimension within each time period. For each $\lambda_{abc}$ there is a corresponding scalar parameter, $\delta_{abc}$, shown in the resulting exposure-lag-response surface at right. To impose monotonicity we require $\delta_{abc}\leq \delta_{abc'}$ for $c<c'$. For identifiability and variable selection we set $\delta_{ab1}=0$.}
    \label{fig:nested_tree_dlnm}
\end{figure}

Our approach to estimating a DLNM with monotone exposure-response and variable selection uses a nested regression tree framework \citep{Mork2021HeterogeneousPollution}. Figure \ref{fig:nested_tree_dlnm} visualizes the nested tree framework. Like most tree methods, we use an ensemble approach. In our case, we use an ensemble of $A$ nested tree units indexed by $a$. Let $\mathcal{T}_a$ denote a binary tree with dichotomous splits on the lagged time periods of exposure $l=0,\ldots,L$ into one or more mutually exclusive segments. The regression tree $\mathcal{T}_a$ consists of internal nodes with splits on the available times and terminal nodes denoted $\{\eta_{ab}\}_{b=1}^{B_a}$ that identify the tree endpoints. In contrast to the previously proposed, unconstrained TDLNM, internal nodes of $\mathcal{T}_a$ split only on time and do not split on values of exposure concentration. Instead, for each terminal node $\{\eta_{ab}\}_{b=1}^{B_a}$ in tree $\mathcal{T}_a$,  we define a binary nested tree $\mathcal{E}_{ab}$. The internal nodes of each $\mathcal{E}_{ab}$ split on values of exposure concentration and the terminal nodes are denoted by $\{\lambda_{abc}\}_{c=1}^{C_{ab}}$. To complete the nested tree model we define a scalar parameter $\delta_{abc}$ corresponding to each $\lambda_{abc}$. Each $\delta_{abc}$ characterizes the exposure-response for a given exposure-concentration and time combination defined by $\mathcal{T}_a$ and $\mathcal{E}_{ab}$.

From the nested tree model, the exposure-lag-response function, $w$, is calculated 
\begin{equation}
  w(x_{t-l},l)=\sum_{a=1}^A \sum_{b=1}^{B_a} \sum_{c=1}^{C_{ab}} \delta_{abc}\psi(x_{t-l},l;\eta_{ab},\lambda_{abc},\sigma_x).
\end{equation}
Here, the sums are over, from left to right: a) nested tree unit in the ensemble, b) terminal node of the time tree $\mathcal{T}_a$, and c) terminal nodes of the nested exposure-splitting tree $\mathcal{E}_{ab}$. The summand contains both the terminal node parameter $\delta_{abc}$ and an exposure-specific weight function denoted by $\psi$ that depends on exposure timing, the terminal nodes and a hyperparameter $\sigma_x$. The weight function $\psi$ can take many forms, two explored by \cite{Mork2022TreedModels} include a step function and a smooth weight function to incorporate smoothness in the exposure-concentration dimension. As most biological applications assume a smooth effect of exposure, we follow the latter approach and define
\begin{equation}
    \psi(x_{t-l}, l; \eta_{ab},\lambda_{abc},\sigma_x)=
    \left[
        \Phi\left(\frac{\lceil x_{abc}\rceil-x_{t-l}}{\sigma_x}\right)-
        \Phi\left(\frac{\lfloor x_{abc}\rfloor-x_{t-l}}{\sigma_x}\right)
    \right] 
    \cdot \mathbb{I}(l\in\eta_{ab}),
\end{equation}
where $\lceil x_{abc}\rceil$ and $\lfloor x_{abc}\rfloor$ are the upper and lower exposure-concentration limits of node $\lambda_{abc}$, respectively, $\Phi$ is the normal cumulative density function, $\mathbb{I}(l\in\eta_{ab})$ is an indicator function that equals 1 when lag time $l$ is in node $\eta_{ab}$ and 0 otherwise, and $\sigma_x$ is a tuning parameter that is fixed for all exposure-time-response functions. Larger $\sigma_x$ will increase the smoothness of the exposure-response curves while smaller $\sigma_x$ allows for sharper changes in the exposure-response relationship. As $\sigma_x\rightarrow 0$ we arrive at the step weight. We treat $\sigma_x$ as fixed due to the computational expense required to estimate it in our model and follow \citet{Mork2022TreedModels} by setting $\sigma_x$ equal to half the standard deviation of the exposure data.

\section{Monotone treed exposure-lag-response function}\label{sec:monotone}
The exposure-response relationship is monotone if $w(x_{t-l},l) \leq w(x_{t-l}^*,l)$ for any two exposures such that $x_{t-l} < x_{t-l}^*$ at the same time $t-l$. We do not assume strict monotonicity as we wish to allow for a null exposure-response relationship. The monotonicity constraint in the exposure-response at each lag time $l$ is satisfied if the terminal node parameters $\{\delta_{abc}\}_{c=1}^{C_{ab}}$ are non-decreasing in the exposure-concentrations spanned by terminal nodes $\{\lambda_{abc}\}_{c=1}^{C_{ab}}$ for each nested tree $\mathcal{E}_{ab}$. Without loss of generality, we order the terminal nodes on each nested tree according to their exposure ranges such that  $\lfloor x_{ab1}\rfloor< \cdots < \lfloor x_{abC_{ab}}\rfloor$, where $\lfloor x_{abc}\rfloor$ is the minimum exposure concentration value in terminal node $\lambda_{abc}$. We then constrain the terminal node parameters $\delta_{ab1}\leq \cdots\leq \delta_{abC_{ab}}$. We set $\delta_{ab1}=0$, which allows for identifiability of the exposure-lag-response function. Specifically, this imposes the identifiability constraint that the exposure-response function at each lag is constrained to be zero at the lowest observed exposure level.

We implement the constraint on the $\delta$'s through a reparameterization of the node-specific parameters for each nested tree following the approach used by \citep{Wang2008a} for monotone regression with Bernstein polynomials. Let $\boldsymbol{\theta}_{ab}=(\theta_{ab1},\dots,\theta_{abC_{ab}})'$ and $\boldsymbol{\delta}_{ab}=(\delta_{ab1},\dots,\delta_{abC_{ab}})'$. We consider first-order difference transformation matrix $\mathbf{D}_{ab}$ such that $\mathbf{D}_{ab}\boldsymbol{\delta}_{ab}=\boldsymbol{\theta}_{ab}$. Specifically, we define the transformation $\delta_{abc}-\delta_{ab(c-1)} = \theta_{abc}$ for $c\geq 2$ and $\theta_{ab1}= \delta_{ab1}=0$. Using this transformation, the parameter space for each $\theta_{abc}$ is fixed to be $\theta_{abc}\geq 0$. This facilitates more efficient sampling via MCMC. In contrast, the parameter space of each $\delta_{abc}$ is constrained such that $\delta_{abc-1}\le \delta_{abc}\le \delta_{abc+1}$ and is challenging to estimate with MCMC. The matrix $\mathbf{D}_{ab}$ follows a block style with adjacent $1$ and $-1$ entries at a unique location in each row and zeros elsewhere. For example, with $C_{ab}=4$, the transformation matrix is
\begin{equation}
\mathbf{D}_{ab}=\left[\begin{array}{rrrr}
 1 &  0 &  0 &  0 \\
-1 &  1 &  0 &  0 \\
 0 & -1 &  1 &  0 \\
 0 &  0 & -1 &  1  \\
\end{array}\right].
\end{equation}
A block-style transformation can be extended to the entire nested tree, such that $\mathbf{D}_a\boldsymbol{\delta}_a=\boldsymbol{\theta}_a$ where $\boldsymbol{\delta}_a=(\boldsymbol{\delta}_{a1}',\ldots, \boldsymbol{\delta}_{aB_a}')'$, $\boldsymbol{\theta}_a=(\boldsymbol{\theta}_{a1}',\ldots, \boldsymbol{\theta}_{aB_a}')'$, and
\begin{equation}
    \mathbf{D}_a=\left[\begin{array}{ccccc}
    \mathbf{D}_{a1} & 0 & \cdots & 0 & 0 \\
    0 & \mathbf{D}_{a2} & \cdots & 0 & 0 \\
    \vdots & \vdots & \ddots & \vdots & \vdots \\
    0 & 0 & \cdots & \mathbf{D}_{a(B_a-1)} & 0 \\
    0 & 0 & \cdots & 0 & \mathbf{D}_{aB_a} \\
\end{array}\right]
\end{equation}
where $\mathbf{D}_{ab}$ make up the diagonal block matrices with zeros elsewhere. We assign independent priors to $\theta_{abc}$, where each follows a truncated normal distribution with range $[0,\infty)$, mean 0, and variance $\sigma^2\nu^2$. In the variance, $\sigma^2$ is the residual variance from a Gaussian model and is fixed to a value of 1 for binomial models. This transformation and prior specification allows us to implement efficient multivariate truncated normal sampling procedures.

%

\section{Zero-inflated regression trees for lag selection}\label{sec:zirt}
The standard BART model uses a splitting probability that decays quickly as the depth of the split increases but assigns a large probability of splitting to the initial node \citep{Chipman2012}. Specifically, BART defines the probability of a split at node $\lambda$ by $p_{\text{split}}(\lambda)=\alpha(1+d_\lambda)^{-\beta},\alpha\in(0,1),\beta>0$, where $d_\lambda$ is the depth of the node beginning at zero. In addition, a probability on the dichotomous splitting rule (i.e., variable and location) is defined, most commonly by a uniform probability across variables and locations. In our situation, we have two sets of trees, the time trees $\mathcal{T}_a$'s and the exposure-response trees $\mathcal{E}_{ab}$'s, that both have different purposes and require separate consideration for an appropriate prior structure.

\subsection{Prior on time-splitting trees \texorpdfstring{$\mathcal{T}_a$}{Ta}}\label{sec:prior_T}

We retain the original BART splitting prior for trees $\mathcal{T}_a$. Imposing default hyperpriors $\alpha_\mathcal{T}=0.95$ and $\beta_\mathcal{T}=2$, results in a prior distribution for the number of terminal nodes in the time dimension with $\mathbb{P}(B_a>1)=0.95$ with $\mathbb{E}(B_a)=2.51$. Having a high prior probability of multiple terminal nodes in the time dimension allows us to identify critical windows and have flexibility across lags. A high prior probability on $B_a$ being relatively small helps to regularize the model by retaining the assumption that the lagged exposure-response functions are similar across nearby times.

For a model with $l=0,\ldots,L$ lags there are $L$ possible splitting points. To determine the time splitting points, we incorporate work by \citet{Linero2018b} where the probability of a rule splitting between lags $l$ and $l+1$ is equal to $\mathcal{P}_{l/l+1}$ with $\mathcal{P}_{0/1},\ldots,\mathcal{P}_{L-1/L}\sim\text{Dirichlet}\{\kappa L^{-1},\ldots,\kappa L^{-1}\}$ and $\kappa (\kappa+L)^{-1}\sim\text{Beta}(1,1)$. This allows for a data-informed approach to identifying change points in the exposure-lag-response function by sharing information about split points across trees in the ensemble.

\subsection{Prior on exposure-response trees \texorpdfstring{$\mathcal{E}_{ab}$}{Eab}}

For the nested trees, $\mathcal{E}_{ab}$, the default BART splitting prior is less desirable because at least two terminal nodes in exposure concentration implies a nonzero effect of exposure. The prior probability of no splits across an ensemble of $A$ trees is $(1-\alpha)^A$. Changing $\alpha$ to reflect prior belief that there is a non-zero effect requires an extremely low $\alpha$ value for even a moderate ensemble size $A$. A negative consequence of setting $\alpha$ to be low is that it will also decrease the probability of splits at all subsequent levels of the nested tree. This is undesirable as it is likely to induce strong shrinkage on the exposure-response relationship. Our objective is, therefore, to allow user-specification of the probability of the first split without impacting the prior probability of subsequent splits conditional on there being a first split. 

We enlist two strategies to help accomplish variable selection in our model. First we define nested trees without splits (i.e. $C_{ab}=1$) to be zero effect (as described by setting $\delta_{ab1}=0$ in Section \ref{sec:monotone}), ensuring that specific time periods will be excluded from the exposure-lag-response function (e.g., these times are not selected). Second, we propose an alternative splitting probability on trees that allows for the time periods with or without effects to also be learned from the data.

For nested tree $\mathcal{E}_{ab}$ we define the zero-inflated tree splitting prior as
\begin{equation}\label{eq:zirt}
    p_{\text{splitZI}}(\lambda|\eta_{ab},\boldsymbol{\gamma},\alpha,\beta)=\pi_0(\boldsymbol{\gamma},\eta_{ab})\mathbb{I}(d_\lambda=0)+\alpha_\mathcal{E}(1+d_\lambda)^{-\beta_\mathcal{E}}\mathbb{I}(d_\lambda>0)
\end{equation}
where $\boldsymbol{\gamma}=[\gamma_0,\ldots,\gamma_L]$ is a vector of parameters corresponding to each lag time point; $\alpha_\mathcal{E}=0.95$ and $\beta_\mathcal{E}=2$ are the standard BART parameters set to typical default values; $d_\lambda$ is the depth of node $\lambda$; and 
\begin{equation}\label{eq:pi0}
    \text{logit}\{\pi_0(\boldsymbol{\gamma},\eta_{ab})\}= |\eta_{ab}|^{-1}\sum_{l\in\eta_{ab}}\gamma_l,
\end{equation}
where $|\eta_{ab}|$ equals the number of times, $l$, contained in node $\eta_{ab}$.
By defining $\pi_0$ with the logistic function, we allow for continuous parameters $\gamma_l\in\mathbb{R}$ and the probability of a split to be based on the time periods in a given terminal node, $\eta_{ab}$. If $\eta_{ab}$ contains many time points with higher valued $\gamma_l$ there will be a high probability of an effect within $\mathcal{E}_{ab}$. If $\eta_{ab}$ instead contains times with many lower valued $\gamma_l$, nested tree $\mathcal{E}_{ab}$ will have a much lower probability of splitting and therefore no effect.

For the zero-inflated splitting parameters, $\gamma_l$, we define prior $\boldsymbol{\gamma} \sim \mathcal{MVN} (\boldsymbol{\gamma}_0,\Sigma)$, where $\boldsymbol{\gamma}_0=(\gamma_{01},\dots,\gamma_{0L})'$ is a prior mean and $\Sigma$ is a covariance matrix. It is natural to set $\gamma_{0l}= 0$ for all $l=0,\dots,L$ implying a prior probability of effect at time $l$ equal to 0.5. However, prior information on which weeks have a nonzero association can be easily incorporated by assigning lag-specific values to each $\gamma_{0l}$. For $\Sigma$ we rely on a scaled identity matrix allowing for the regularization of the tree structures to impose a model-specified correlation among $\boldsymbol{\gamma}$. For instance, in our simulation as a default prior we set $\boldsymbol{\gamma}_0= \mathbf{0}$ and $\Sigma=0.314\cdot\mathbf{I}_L$, where $\mathbf{I}_L$ denotes an $L\times L$ identity matrix, will define a prior such that 95\% of the time $\pi_0(\gamma_{0l})$ falls between 0.25 and 0.75 for each time, $l$. In our data analysis we incorporate an even more vague prior by setting $\boldsymbol{\gamma}_0= \mathbf{0}$ and $\Sigma=7.294\cdot\mathbf{I}_L$ which says that 95\% of the time $\pi_0(\gamma_{0l})$ will fall between 0.005 and 0.995.


\subsection{Variable-selection based inference on lags}
In the spline-based DLNM or TDLNM, periods of susceptibility (i.e., lags with nonzero exposure-response relationships) are not well defined. They are typically identified as lag times that have credible intervals not containing zero for some user-specified exposure contrast of interest. In the proposed method, we can directly infer the probability of a lag-specific susceptibility through a posterior analysis of terminal nodes at a given lag time. Specifically, for posterior samples $r=1,\ldots, R$ define $E_l^{(r)}=1$ if any nested tree $\mathcal{E}_{ab}$ with $l\in\eta_{ab}$ has 2 or more terminal nodes, otherwise $E_l^{(r)}=0$. Then,
\begin{equation}
    \hat{\mathbb{P}}(\text{susceptibility at lag $l$})= R^{-1}\sum_{r=1}^R E_l^{(r)}.
\end{equation}
By specifying a reasonable level of confidence (e.g., probability $\geq 0.95$) we can make a conclusion about which lag times show susceptibility to the exposure. It is noted that at $\hat{\mathbb{P}}(\text{susceptibility at lag }l)=0.95$, the corresponding central credible interval will equal zero as the lower bound is at the $2.5$ percentile. If this result is not desired, alternative solutions are to use an upper 95\% credible interval or a 90\% credible interval alongside the probability of effect.

\section{Prior specification and posterior computation}
\subsection{Prior specification}

For simplicity, we focus on a Gaussian model in this section. The model is
\begin{equation}
    y_t=\sum_{l=0}^L w(x_{t-l},l) + h(t;\boldsymbol{\zeta}) + \varepsilon_t,
\end{equation}
where $\varepsilon_t\sim\mathcal{N}(0,\sigma^2)$ and assumed independent. We discuss binomial outcomes in Section~\ref{sec:binary}. 

We complete the Bayesian specification of the model by assigning priors to the remaining parameters. We specify half-Cauchy priors for the variance parameters, with $\sigma,\nu\sim\mathcal{C}^+(0,1)$. Because $\sigma^2$ is the variance parameter for the residuals in the fixed effect model, it can be interpreted as a scaling factor on residual variance in the tree ensemble, $\nu^2$, and is equivalent to scaled BART variance prior described by \citet{Linero2018}. We specify a non-informative prior on the regression parameters for the time trend and covariate model, $\boldsymbol\zeta\sim\mathcal{MVN}(\mathbf{0},c\sigma^2\mathbf{I})$, where $c$ is fixed to a large value.

\subsection{MCMC approach for Gaussian model}

We estimate the model parameters using MCMC with a hybrid Gibbs-Metropolis-Hastings algorithm. Specifically, the terminal node parameters can be efficiently sampled via a Gibbs sampler as can the variance components and regression parameters for the covariates. The tree structures are updated with the Metropolis-Hastings (MH) algorithm using the grow, prune, and change proposals steps. When updating the exposure-splitting trees we also consider the zero-inflation probability and update that probability. 

\subsubsection{Updating terminal node parameters: \texorpdfstring{$\boldsymbol\theta_a$}{thetaa}}
Define the total exposure effect as $f(\mathbf{x}_t)=\sum_{l=0}^L w(x_{t-l},l)$. Let $u_{tabc}=\sum_{l=0}^L\mathbb{I}(x_{t-l} \in\lambda_{abc}, l\in\eta_{ab})$ be the count of lagged exposure measurements prior to $t$ within the exposure concentration range of terminal node $\lambda_{abc}$ and the time range of terminal node $\eta_{ab}$. Define $\mathbf{u}_{ta}=(u_{ta11}, \dots, u_{ta1C_{a1}}, u_{ta21}, \dots, u_{taB_{a}C_{aB_a}})'$ and $\mathbf{U}_a$ be the matrix containing rows $\mathbf{u}_{ta}'$. Applying the transformation from Section \ref{sec:monotone}, $\mathbf{u}_{ta}'\boldsymbol\delta_a=\mathbf{u}_{ta}'\mathbf{D}_a^{-1}\boldsymbol\theta_a$ and the total exposure effect is $f(\mathbf{x}_t)=\sum_{a=1}^A\mathbf{u}_{ta}'\mathbf{D}_a^{-1}\boldsymbol\theta_a$.

To estimate the exposure-lag-response function we integrate out $\boldsymbol\zeta$ from the full likelihood. Let $\mathbf{y}=(y_1,\ldots,y_n)'$ and $\mathbf{f}=[f(\mathbf{x}_1),\ldots,f(\mathbf{x}_n)]'$ be vectors of length $n$. Then,
\begin{align*}
    \mathbf{y}|\sigma &\sim\mathcal{MVN}(\mathbf{f},\sigma^2\mathbf{V}_\mathbf{Z})\\
    \mathbf{V}_\mathbf{Z} &=(\mathbf{I}-\mathbf{Z}'\mathbf{V}_{\boldsymbol\zeta}\mathbf{Z})^{-1}\\
    \mathbf{V}_{\boldsymbol\zeta} &=(\mathbf{Z}'\mathbf{Z}+c^{-1}\mathbf{I})^{-1},
\end{align*}
where $\mathbf{Z}$ is a matrix with rows $\mathbf{z}_t'$ that are vector functions of the covariates or spline basis expansion for time. Using a Bayesian backfitting approach \citep{hastie2000bayesian} we consider the partial residuals $\mathbf{R}_a=[R_{1a},\ldots,R_{na}]'$ where $R_{ta}=y_t-\sum_{a'\not=a}\mathbf{u}_{ta'}\mathbf{D}_{a'}^{-1}\boldsymbol\theta_{a'}$ resulting in
\begin{equation*}
    \mathbf{R}_a|\mathcal{T}_a,\mathcal{E}_{a1},\ldots,\mathcal{E}_{aB_a},\boldsymbol\theta_a,\sigma\sim\mathcal{MVN}(\mathbf{U}_a\mathbf{D}_a^{-1}\boldsymbol\theta_a,\sigma^2\mathbf{V}_\mathbf{Z}).
\end{equation*}
The transformed terminal node parameters $\boldsymbol\theta_a$ for tree $\mathcal{T}_a$ and nested trees $\{\mathcal{E}_{ab}\}_{b=1}^{B_a}$ are simulated as a block from their full conditional
\begin{align*}
    \boldsymbol\theta_a|-&\sim\mathcal{TN}_{[\mathbf{0},\boldsymbol\infty)}[\mathbf{V}_{\boldsymbol\theta_a}(\mathbf{U}_a\mathbf{D}_a^{-1})'\mathbf{V}_\mathbf{Z}^{-1}\mathbf{R}_a,\sigma^2\mathbf{V}_{\boldsymbol\theta_a}]\\
    \mathbf{V}_{\boldsymbol\theta_a}&=\left[(\mathbf{U}_a\mathbf{D}_a^{-1})'\mathbf{V}_\mathbf{Z}^{-1}\mathbf{U}_a\mathbf{D}_a^{-1}+\nu^{-2}\mathbf{I}\right]^{-1}.
\end{align*}
Draws for $\boldsymbol\theta_a$ are done via efficient sampling methods proposed by \citet{Li2015EfficientConstraints}.

\subsubsection{Updating \texorpdfstring{$\mathcal{E}_{ab}$}{Eab}}
To facilitate convergence by allowing larger proposal steps in the Markov chain we deviate from the typical BART proposal framework when updating each $\mathcal{E}_{ab}$ and instead draw directly from the tree prior. 
Recall that the probability of nested tree $\mathcal{E}_{ab}$ equals
\begin{equation*}
    p(\mathcal{E}_{ab})=\prod_{\lambda\; \text{internal}}
    p_{\text{splitZI}}(\lambda|\eta_{ab},\boldsymbol{\gamma},\alpha_\mathcal{E},\beta_\mathcal{E})p_{\text{rule}}(\lambda)
    \prod_{\lambda\; \text{terminal}}\left[1-p_{\text{splitZI}}(\lambda|\eta_{ab},\boldsymbol{\gamma},\alpha_\mathcal{E},\beta_\mathcal{E})\right].
\end{equation*}
Our algorithm for drawing proposed nested tree $\mathcal{E}^*$ starts by drawing a new tree from the prior, without any consideration for the observed data. First, $\mathcal{E}_{ab}^*$ splits from root with probability $\pi_0(\boldsymbol{\gamma},\eta_{ab})$. If a split from root occurs, this results in two new terminal nodes, $\lambda_{ab1}$ and $\lambda_{ab2}$. Each terminal node, $\lambda_{abc}$, splits with probability $\alpha_\mathcal{E}(1+d_{\lambda_{abc}})^{\beta_\mathcal{E}}$, resulting in two new terminal nodes (the splitting node is now considered an internal node). This process continues with each new terminal node until no new terminal nodes are created. Next, accept or reject the proposed tree drawn from the prior using the observed data and MH. 
The MH algorithm acceptance probability for proposed nested tree $\mathcal{E}_{ab}^*$ 
equals
\begin{eqnarray*}
r&=&\min\left\{
    \frac{
        p(\mathbf{R}_a|\mathcal{T}_a,\{\mathcal{E}_{ab}\}^*,\nu,\sigma) p(\mathcal{E}_{ab}^*)p(\mathcal{E}_{ab}|\mathcal{E}_{ab}^*)
    }{
        p(\mathbf{R}_a|\mathcal{T}_a,\{\mathcal{E}_{ab}\},\nu,\sigma)  p(\mathcal{E}_{ab})p(\mathcal{E}_{ab}^*|\mathcal{E}_{ab})
    },1
\right\}
\end{eqnarray*}
where $\{\mathcal{E}_{ab}\}=\mathcal{E}_{a1},\ldots,\mathcal{E}_{aB_a}$ is the set of nested trees and $\{\mathcal{E}_{ab}\}^*=\mathcal{E}_{a1},\ldots,\mathcal{E}_{ab}^*,\ldots,\mathcal{E}_{aB_a}$ is the set of nested trees with a tree proposal. Here, the proposal distribution, $p(\mathcal{E}_{ab}^*|\mathcal{E}_{ab})=p(\mathcal{E}_{ab}^*)$ requiring only that we evaluate the likelihood, $p(\mathbf{R}_a|\mathcal{T}_a,\{\mathcal{E}_{ab}\},\nu,\sigma)$. 
We calculate
\begin{align*}
p(\mathbf{R}_a|\mathcal{T}_a,\{\mathcal{E}_{ab}\},\nu,\sigma) =&
\int p(\mathbf{R}_a|\mathcal{T}_a,\{\mathcal{E}_{ab}\},\boldsymbol\theta_a,\sigma)p(\boldsymbol\theta_a|\sigma,\nu)d\boldsymbol\theta_a\\
=&\;(2\pi\sigma^2)^{-n/2}(2^2\sigma^2\nu^2)^{-p/2}|\mathbf{V}_\mathbf{Z}|^{-1/2}|\mathbf{V}_{\boldsymbol\theta_a}|^{1/2}\\
&\times\exp\left\{-\frac{\mathbf{R}_a'\mathbf{V}_\mathbf{Z}^{-1}\mathbf{R}_a}{2\sigma^2}+\frac{\mathbf{R}_a'\mathbf{V}_\mathbf{Z}^{-1}\mathbf{U}_a\mathbf{D}_a^{-1}\mathbf{V}_{\boldsymbol\theta_a}(\mathbf{U}_a\mathbf{D}_a^{-1})'\mathbf{V}_\mathbf{Z}^{-1}\mathbf{R}_a}{2\sigma^2}\right\}\\
&\times \left[\int_{[\mathbf{0},\boldsymbol\infty)}p(\boldsymbol\theta_a|\mathbf{R}_a,-)d\boldsymbol\theta_a\right],
\end{align*}
where the last integral is the normalizing constant for the full conditional distribution of $\boldsymbol\theta_a$, which can be numerically approximated through algorithms described by \citet{Genz2012ComparisonProbabilities}. 

\subsubsection{Updating \texorpdfstring{$\mathcal{T}_a$}{Ta}}
In our nested tree framework an additional consideration must be made when updating tree $\mathcal{T}_a$. When we grow $\mathcal{T}_a$ we must replace a nested tree, $\mathcal{E}_{ab}$ with two new nested trees, $\mathcal{E}_{ab_1}^*$ and $\mathcal{E}_{ab_2}^*$. Therefore, proposed tree $\mathcal{T}_a^*$ has a different number of terminal nodes, $B_a^*$, each with a corresponding nested tree. When we apply a grow proposal $\mathcal{T}_a^*$, we draw each new $\mathcal{E}_{ab_i}^*$, $i\in\{1,2\}$, using the the previously described algorithm. The other nested trees remain the same. To calculate the MH acceptance ratio for proposal $\mathcal{T}_a$ we follow similar steps to updating $\mathcal{E}_{ab}$ by first calculating $p(\mathbf{R}_a|\mathcal{T}_a^*,\{\mathcal{E}_{ab}\}^*,\nu,\sigma)$ and $p(\mathbf{R}_a|\mathcal{T}_a,\{\mathcal{E}_{ab}\},\nu,\sigma)$. The resulting acceptance ratio equals
\begin{equation}
    r=\min\left\{\frac{p(\mathbf{R}_a|\mathcal{T}_a^*,\{\mathcal{E}_{ab}\}^*,\nu,\sigma)p(\mathcal{T}_a^*)p(\mathcal{T}_a|\mathcal{T}_a^*)}{p(\mathbf{R}_a|\mathcal{T}_a,\{\mathcal{E}_{ab}\},\nu,\sigma)p(\mathcal{T}_a)p(\mathcal{T}_a^*|\mathcal{T}_a)},1\right\}.
\end{equation}
A prune proposal for $\mathcal{T}_a$ replaces two nested trees corresponding to the pruned nodes with a new nested tree while a change proposal is the same as changing an internal node in standard BART. In a change proposal for $\mathcal{T}_a$ we retain the structure of all nested trees.

\subsubsection{Updating remaining parameters: \texorpdfstring{$\gamma_l,\boldsymbol\zeta$, $\sigma, \text{and } \nu$}{gamma, zeta, sigma, and nu}}
Each terminal node $\eta_{ab}$ contributes one binary `observation' (effect or no effect) to a logistic model relating the probability of an effect at lag time $l$ and parameters $\gamma_l$. We employ a Poly\'a-gamma augmentation approach to updating $\gamma_l$ \citep{Polson2013BayesianVariables}.

The remaining parameters are updated with Gibbs steps using standard full conditionals. The variance components $\sigma$, and $\nu$ are updated using the approach of \cite{Makalic2016AEstimator}. The parameters for the time trend and covariate model, $\boldsymbol\gamma$ and $\boldsymbol\zeta$, have multivariate normal full conditionals.

\subsection{Implementation for binomial response}\label{sec:binary}
For a binomial response, $y_t$, we define the logistic model,
\begin{equation}
    y_t|\mathbf{x}_t,t\sim\text{Binomial}\{n_t,1/(1+e^{-\psi_t})\}
\end{equation}
where $\psi_t=\sum_{l=0}^L w(x_{t-l},l) + h(t;\boldsymbol{\zeta})$.
To estimate the exposure-lag-response function, $w(x_{t-l},l)$, we follow a Poly\'a-gamma augmented variable approach. Briefly, for Poly\'a-gamma random variable, $\omega_t\sim\text{PG}(n_t,\psi_t)$, the data-likelihood is proportional to $\exp\{(y_t-n_t/2)\psi_t\}\mathbb{E}\{\exp(-\omega_t\psi_t^2/2)\}$, where the expectation is with respect to a PG$(n_t,0)$ random variable (see e.g. \citet{Polson2013BayesianVariables}). The inclusion of $\omega_t$ creates a conditionally Gaussian likelihood and allows for Gibbs updates of $\boldsymbol\theta_{a}$ and $\boldsymbol\zeta$ parameters.

\subsection{Strategies for incorporating additional prior information}\label{sec:inform_prior}
Two goals that often exist simultaneously in estimating exposure-lag-response functions are identifying periods of susceptibility or nonzero effects in the exposure response and change points in the lag response dimension. The specification of monotone-TDLNM allows for the inclusion of prior information in each dimension. 

To increase or decrease the prior probability of a nonzero exposure relationship we can alter the priors in the zero-inflated regression tree for specific time periods, specifically $\boldsymbol\gamma_0$ and $\Sigma_p$. For example, we can define $\gamma_{0l}$ and $\Sigma_p(l,l)$ such that $1/\{1+e^{-\gamma_{0l}}\}$ falls within a given range 95\% of the time. In practice, high probability of effect in some lags should be balanced by lower probability of effect in other lags to maintain roughly zero mean across $\boldsymbol\gamma_0$ to prevent false positives when many lag periods are included in the same terminal node.

Informative priors for lag change points are easily introduced via two alternate approaches. First, we can fix the time split probability rules, $\mathcal{P}_{l/l+1}$, to increase the probability of a tree splitting at a given lag and increasing the likelihood that different exposure-response relationships will exist before or after the split. In some analyses where there is a well-defined change point in the exposure time-series, this approach makes the most sense. For example, air pollution exposures during fetal development and early childhood may assume a change in the exposure-lag-response at the time of birth--here it makes sense to apply a large fixed probability of split at that time point. The second strategy to incorporating prior information into the time split probabilities is through a modification to the Dirichlet prior. Here, we introduce prior $\mathcal{P}_{0/1},\ldots,\mathcal{P}_{L-1/L}\sim\text{Dirichlet}(d_{0/1}\kappa,\ldots,d_{L-1/L}\kappa)$ where $\sum d_{l/l+1}=1$, and $d_{l/l+1}>0$ is the prior probability of a split point between lag $l$ and $l+1$ in a given tree. The hyper parameter $\kappa$ may be set for a specific variance in the probabilities or assigned a prior distribution as in Section \ref{sec:prior_T}. Using a fixed $\kappa$ in the Dirichlet prior can allow for additional posterior inference on the change points through Bayes Factors or similar methods.

An alternative method to introduce prior information into the regression tree model is using fixed time trees corresponding to previously estimated periods of susceptibility. For example, if two other studies identify periods of susceptibility during specific lag periods, we may fix two trees in our model with change points corresponding to these past studies. The model and data will determine whether the exposure-lag-response function in the nested trees is nonzero; however, we remove the need to select splitting times in these trees. Other trees in the model can continue to explore other splitting times in case the relationships in the data do not agree with previously estimated lag periods of susceptibility. Furthermore, the idea of fixed time trees could allow for information transfer from larger studies, where the time periods can be more effectively estimated, to small studies. The idea of using one sample to estimate the periods of susceptibility and another sample to estimate the exposure response is closely tied to the idea of honest inference \citep{wager2018estimation, athey2016recursive}.

\section{Simulation study}
We developed a simulation study based on our data analysis to compare our proposed method alongside existing methods TDLNM \citep{Mork2022TreedModels} and the penalized spline DLNM (GAM) \citep{Gasparrini2017}. The goals of our simulation study were to first, show that the monotonicity assumptions in monotone-TDLNM improve estimation of the exposure-lag-response function while providing valid inference in terms of confidence intervals and high precision for identifying periods of susceptibility and second, that the inclusion of informative priors can additionally supplement ability of DLNM methods to estimate exposure-lag-response functions.

\begin{figure}[!ht]
     \centering
     \begin{subfigure}[t]{6cm}
         \centering
         \includegraphics[width=6cm]{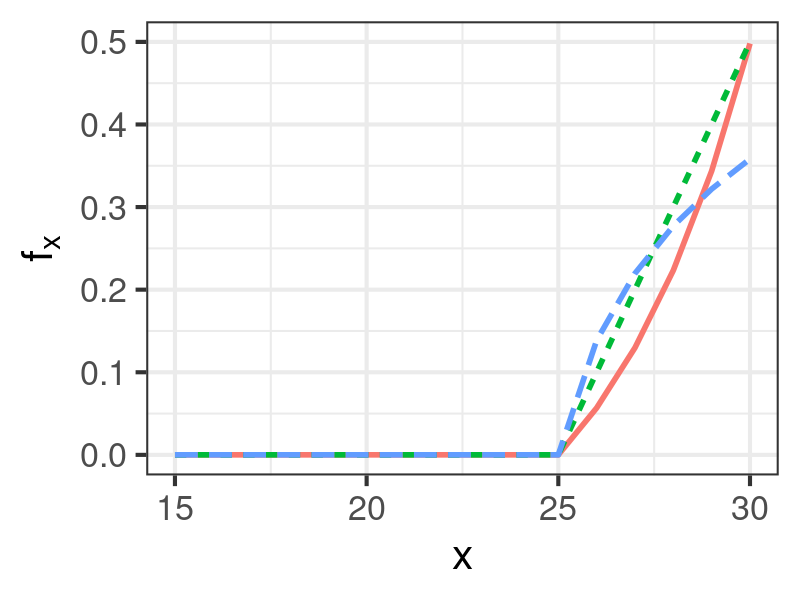}
     \end{subfigure}
     \begin{subfigure}[t]{6cm}
         \centering
         \includegraphics[width=6cm]{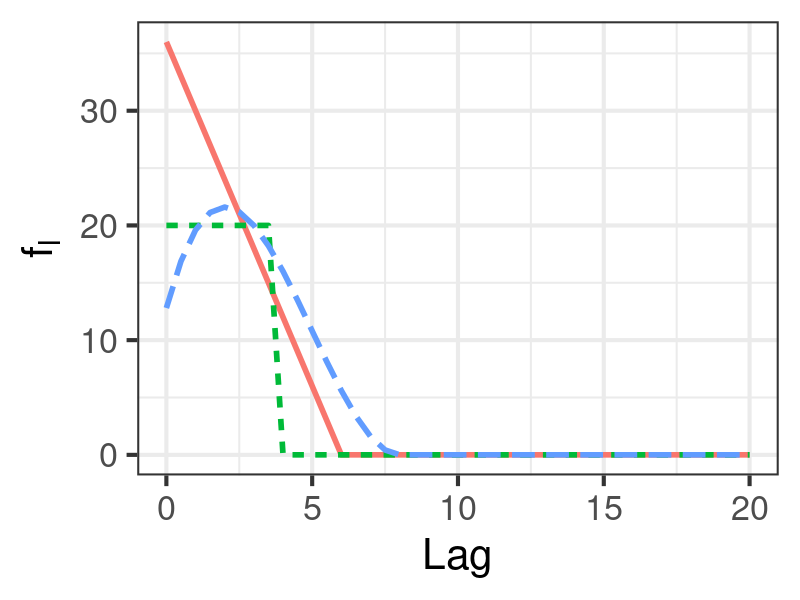}
     \end{subfigure}
     \caption{The exposure-lag-response function in our simulations is defined by $w(x_{t-l},l)= f_x(x_{t-l})\cdot f_l(l)$, based on combinations of $f_x$ and $f_l$.}
     \label{fig:sim_dlnm}
\end{figure}

For each simulation replicate we sampled $n=1000$ days from the summer temperature time series in our data analysis and used $L=20$ days of lagged temperature exposures. We created 9 simulation scenarios based on combinations of 3 exposure-response relationships given by
\begin{equation}
    f_x(x) = \mathbb{I}(x > 25)
    \begin{cases}
        0.1\cdot(x-25) & \text{linear}\\
        0.2\cdot\log(x - 25) & \text{sublinear}\\
        0.2\cdot[\exp\{0.25\cdot(x-25)\}-1] & \text{exponential}
    \end{cases}
\end{equation}
and 3 lag-response relationships,
\begin{equation}
    f_l(l) = 
    \begin{cases}
        20\cdot\mathbb{I}(l < 4) & \text{piecewise}\\
        \max\{0,6\cdot (6 - l)\} & \text{linear}\\
        \max\{0, 0.2\cdot (l+1)\cdot (l-8)^2\} & \text{quadratic}
    \end{cases}
\end{equation}
where the exposure-lag-response function is calculated as $w(x_{t-l},l)=f_x(x_{t-l})\cdot f_l(l)$. Figure \ref{fig:sim_dlnm} shows plots of $f_x$ and $f_l$. When considering the exposure-response and lag-response functions separately, the piecewise lag-response function can be exactly fit with a single tree with two terminal nodes. All other exposure-response and lag-response functions require multiple splits from a single tree, or more realistically an ensemble of trees, to approximate the function. When the exposure-response and lag-response functions are combined, all nine scenarios require  complex tree structures to approximate the exposure-lag-response function.

The outcome was simulated by $y_t=\sum_{l=0}^{20}w(x_{t-l},l)+ \varepsilon_t$ where $\varepsilon_t$ was drawn independently from a normal distribution with mean zero and standard deviation 2, 4, and 8 times the standard deviation of the exposure-lag-response. We repeated each combination of exposure-response, lag-response, and error for 100 simulation replicates.

For each simulation scenario, we compare three models: GAM, TDLNM, and monotone-TDLNM. We also compare with the same models using informative priors. The objective of adding informative priors is not simply to get better results but to provide empirical evidence documenting the effects of including this form of correct prior information into an analysis with the proposed models. For GAM, we use a penalized B-spline crossbasis with 10 degrees of freedom in both exposure and lag dimensions. To incorporate informative priors, we add varying ridge penalization to the latter 6 lag degrees of freedom which will increase the shrinkage of estimates in later lag periods toward zero  (see e.g., \citet{Gasparrini2017} for more details). We apply TDLNM and monotone-TDLNM with default priors as described in \citet{Mork2022TreedModels} and this paper. To add informative prior information to TDLNM we define the probability of a split in time to be 10 times higher during the lag periods of true effect compared to other lag periods. For monotone-TDLNM, we add informative priors in two places as described in Section \ref{sec:inform_prior}: first we set the mean and variance of the prior normal distribution for $\gamma_l$ such that there is a 95\% probability $\pi_0(\gamma_l)$ is between 0.95 and 0.995 during lag periods of effect and between 0.005 and 0.995 during other lag periods; second we set splitting parameters such that $d_{l/l+1}$ divided by $d_{l'/l'+1}$ equals 10 for $l$ and $l'$ being periods of effect or no effect, respectively; $\kappa$ retains the prior distribution described in Section \ref{sec:prior_T}. For TDLNM and monotone-TDLNM we fix the smoothing parameter $\sigma_x$ equal to half the standard deviation of the temperature data. Because a smooth exposure-response function at each lag enforces strict monotonicity, we add a small quantity of 0.05 to each side of the estimated credible intervals to capture places where the exposure-lag-response function remains at zero across a range of exposure-concentration values (e.g., less than 25 for periods of nonzero effect). For the tree model we combine results from two independent Markov chains. We run each chain for 7,000 total iterations, discarding the first 2,000 iterations as burn-in and retaining every 10th iteration to allow for adequate mixing. Convergence is assessed using the Gelman-Rubin statistic \citep{Gelman1992InferenceSequences}.

For each simulation scenario, we estimate the exposure-lag-response function across a grid of values spanning the temperature and lag data: $x=3,4,5\ldots,30$ and $l=0,\ldots,20$. We calculate the root mean square error (RMSE), first averaged at each point on the grid then averaged across the entire exposure-lag-response function. We determine the average coverage based on how often 95\% credible intervals cover the true exposure-lag-response value at each point on the grid. We also calculate the average credible interval width by taking the average difference between the upper and lower interval limits. Finally, we calculate precision as the proportion of correctly identified time periods of nonzero effect (true positive) relative to the total identified time periods of nonzero effect (true positive plus false positive). In TDLNM and GAM we use credible intervals to determine precision, in monotone-TDLNM we consider $\hat{\mathbb{P}}(\text{susceptibility at time } l)\geq 0.95$ as the criteria for a nonzero effect.

Simulation results for $\sigma=2,8$ containing RMSE, coverage, interval width, and precision are presented in Tables \ref{tab:sigma2} and \ref{tab:sigma8}. Additional results are provided in Supplementary Materials \citep{MorkBASupp}. Monotone-TDLNM and TDLNM consistently have the lowest RMSE across all simulation scenario and error combinations. Targeted penalization decreases the RMSE for the GAM approach, but this method consistently has the highest RMSE. Both TDLNM and monotone-TDLNM are able to shrink estimates towards zero in places of no effect while GAM often overgeneralizes periods of effect and retains additional wiggliness across the estimated exposure-lag-response function. Additionally, the spline based GAM displays more extreme behavior at the boundaries, a characteristic not shared by the tree-based methods.

\begin{table}[!ht]
    \caption{Simulation results for $\sigma=2$, including root mean square error (RMSE), coverage by 95\% credible or confidence intervals, average credible/confidence interval width (upper minus lower) and precision. Results shown for all exposure-response ($f_x$) and lag-response ($f_l$) combinations.}
    \label{tab:sigma2}
\footnotesize
\begin{tabular}{lllccccccc}
\toprule[1pt]
&&&\multicolumn{3}{c}{Standard Prior}&&\multicolumn{3}{c}{Informative Prior}\\
\cmidrule{4-6}\cmidrule{8-10}
Metric & $f_x$ & $f_l$ & GAM & TDLNM & Monotone && GAM & TDLNM & Monotone\\
\midrule
RMSE & Exp & Lin & 1.10 & 0.48 & 0.61 && 0.80 & 0.49 & 0.72\\
 &  & Pw & 0.86 & 0.27 & 0.26 && 0.72 & 0.25 & 0.41\\
 &  & Quad & 0.89 & 0.39 & 0.43 && 0.71 & 0.38 & 0.56\\
 & Lin & Lin & 1.05 & 0.55 & 0.60 && 0.70 & 0.58 & 0.68\\
 &  & Pw & 0.83 & 0.34 & 0.35 && 0.67 & 0.32 & 0.44\\
 &  & Quad & 0.86 & 0.43 & 0.49 && 0.63 & 0.44 & 0.58\\
 & Sub & Lin & 0.94 & 0.48 & 0.53 && 0.60 & 0.50 & 0.60\\
 &  & Pw & 0.71 & 0.30 & 0.27 && 0.56 & 0.27 & 0.39\\
 &  & Quad & 0.74 & 0.38 & 0.40 && 0.57 & 0.38 & 0.49\\
\addlinespace
Coverage & Exp & Lin & 0.96 & 0.99 & 0.98 && 0.99 & 0.99 & 0.98\\
 &  & Pw & 0.95 & 0.99 & 0.98 && 0.98 & 0.99 & 0.99\\
 &  & Quad & 0.94 & 0.98 & 0.97 && 0.98 & 0.98 & 0.97\\
 & Lin & Lin & 0.95 & 0.99 & 0.98 && 0.98 & 0.99 & 0.97\\
 &  & Pw & 0.94 & 0.99 & 0.98 && 0.96 & 0.99 & 0.98\\
 &  & Quad & 0.93 & 0.98 & 0.97 && 0.96 & 0.97 & 0.96\\
 & Sub & Lin & 0.94 & 0.98 & 0.97 && 0.97 & 0.98 & 0.96\\
 &  & Pw & 0.92 & 0.98 & 0.97 && 0.95 & 0.98 & 0.97\\
 &  & Quad & 0.91 & 0.97 & 0.94 && 0.94 & 0.96 & 0.94\\
\addlinespace
CI Width & Exp & Lin & 3.55 & 1.80 & 0.97 && 2.11 & 1.83 & 0.98\\
 &  & Pw & 2.51 & 1.15 & 0.42 && 1.57 & 1.06 & 0.50\\
 &  & Quad & 2.63 & 1.42 & 0.61 && 1.76 & 1.27 & 0.77\\
 & Lin & Lin & 3.35 & 2.19 & 1.05 && 1.92 & 2.01 & 1.20\\
 &  & Pw & 2.38 & 1.34 & 0.55 && 1.38 & 1.28 & 0.56\\
 &  & Quad & 2.58 & 1.64 & 0.85 && 1.51 & 1.54 & 0.81\\
 & Sub & Lin & 2.92 & 1.86 & 0.92 && 1.58 & 1.67 & 0.98\\
 &  & Pw & 2.06 & 1.17 & 0.38 && 1.12 & 0.98 & 0.44\\
 &  & Quad & 2.28 & 1.42 & 0.69 && 1.17 & 1.22 & 0.66\\
\addlinespace
Precision & Exp & Lin & 0.74 & 1.00 & 1.00 && 0.95 & 1.00 & 1.00\\
 &  & Pw & 0.73 & 1.00 & 1.00 && 0.89 & 1.00 & 1.00\\
 &  & Quad & 0.68 & 0.97 & 1.00 && 0.92 & 0.92 & 1.00\\
 & Lin & Lin & 0.73 & 0.97 & 1.00 && 0.88 & 0.89 & 1.00\\
 &  & Pw & 0.73 & 1.00 & 1.00 && 0.88 & 0.95 & 1.00\\
 &  & Quad & 0.68 & 1.00 & 1.00 && 0.88 & 0.90 & 1.00\\
 & Sub & Lin & 0.69 & 0.97 & 1.00 && 0.87 & 1.00 & 1.00\\
 &  & Pw & 0.70 & 1.00 & 1.00 && 0.83 & 1.00 & 1.00\\
 &  & Quad & 0.65 & 0.95 & 1.00 && 0.84 & 0.89 & 1.00\\
\bottomrule[1pt]
\end{tabular}
\end{table}

\begin{table}[!ht]
    \caption{Simulation results for $\sigma=8$, including root mean square error (RMSE), coverage by 95\% credible or confidence intervals, average credible/confidence interval width (upper minus lower) and precision. Results shown for all exposure-response ($f_x$) and lag-response ($f_l$) combinations.}
    \label{tab:sigma8}
\footnotesize
\begin{tabular}{lllccccccc}
\toprule[1pt]
&&&\multicolumn{3}{c}{Standard Prior}&&\multicolumn{3}{c}{Informative Prior}\\
\cmidrule{4-6}\cmidrule{8-10}
Metric & $f_x$ & $f_l$ & GAM & TDLNM & Monotone && GAM & TDLNM & Monotone\\
\midrule
RMSE & Exp & Lin & 2.28 & 1.36 & 1.24 && 1.95 & 1.32 & 1.34\\
 &  & Pw & 1.58 & 0.96 & 0.88 && 1.28 & 0.94 & 0.97\\
 &  & Quad & 1.78 & 1.01 & 0.94 && 1.48 & 1.01 & 1.03\\
 & Lin & Lin & 2.29 & 1.48 & 1.34 && 1.79 & 1.42 & 1.44\\
 &  & Pw & 1.57 & 1.05 & 0.94 && 1.27 & 1.01 & 1.04\\
 &  & Quad & 1.75 & 1.10 & 1.05 && 1.41 & 1.09 & 1.12\\
 & Sub & Lin & 1.91 & 1.22 & 1.12 && 1.44 & 1.19 & 1.19\\
 &  & Pw & 1.34 & 0.87 & 0.83 && 1.06 & 0.85 & 0.85\\
 &  & Quad & 1.48 & 0.92 & 0.87 && 1.17 & 0.91 & 0.93\\
\addlinespace
Coverage & Exp & Lin & 0.94 & 0.98 & 0.99 && 0.94 & 0.98 & 0.95\\
 &  & Pw & 0.93 & 0.98 & 0.99 && 0.94 & 0.99 & 0.98\\
 &  & Quad & 0.93 & 0.97 & 0.99 && 0.93 & 0.98 & 0.90\\
 & Lin & Lin & 0.94 & 0.97 & 0.99 && 0.93 & 0.98 & 0.93\\
 &  & Pw & 0.91 & 0.98 & 0.99 && 0.92 & 0.98 & 0.97\\
 &  & Quad & 0.9 & 0.97 & 0.98 && 0.93 & 0.98 & 0.89\\
 & Sub & Lin & 0.9 & 0.97 & 0.98 && 0.93 & 0.98 & 0.93\\
 &  & Pw & 0.9 & 0.98 & 0.99 && 0.92 & 0.98 & 0.96\\
 &  & Quad & 0.9 & 0.97 & 0.98 && 0.93 & 0.97 & 0.89\\
\addlinespace
CI Width & Exp & Lin & 6.55 & 4.98 & 3.64 && 3.80 & 4.68 & 1.96\\
 &  & Pw & 4.35 & 3.11 & 1.93 && 2.63 & 3.09 & 1.56\\
 &  & Quad & 5.03 & 3.92 & 2.48 && 3.38 & 3.75 & 1.44\\
 & Lin & Lin & 6.32 & 5.14 & 3.28 && 3.61 & 5.16 & 2.08\\
 &  & Pw & 3.71 & 3.36 & 2.16 && 2.73 & 3.35 & 1.63\\
 &  & Quad & 4.40 & 4.29 & 2.58 && 3.45 & 4.21 & 1.54\\
 & Sub & Lin & 4.57 & 4.32 & 2.69 && 2.92 & 4.23 & 1.78\\
 &  & Pw & 3.06 & 2.83 & 1.76 && 2.31 & 3.02 & 1.29\\
 &  & Quad & 3.82 & 3.47 & 2.15 && 2.87 & 3.49 & 1.34\\
\addlinespace
Precision & Exp & Lin & 0.73 & 0.97 & 1.00 && 0.84 & 1.00 & 1.00\\
 &  & Pw & 0.71 & 0.93 & 1.00 && 0.83 & 1.00 & 1.00\\
 &  & Quad & 0.66 & 0.97 & 1.00 && 0.73 & 1.00 & 1.00\\
 & Lin & Lin & 0.73 & 0.92 & 1.00 && 0.74 & 1.00 & 1.00\\
 &  & Pw & 0.70 & 0.96 & 1.00 && 0.73 & 1.00 & 1.00\\
 &  & Quad & 0.63 & 0.97 & 1.00 && 0.72 & 1.00 & 1.00\\
 & Sub & Lin & 0.71 & 0.95 & 1.00 && 0.74 & 1.00 & 1.00\\
 &  & Pw & 0.68 & 0.92 & 1.00 && 0.71 & 1.00 & 1.00\\
 &  & Quad & 0.69 & 0.97 & 0.91 && 0.69 & 1.00 & 1.00\\
\bottomrule[1pt]
\end{tabular}
\end{table}

Monotone-TDLNM generally maintains nominal coverage of the true exposure-lag-response function by 95\% credible intervals and has the smallest credible interval width across the range of simulation scenarios and error settings. Adding informative prior information further decreases the credible interval width in the highest error setting. We note that the strict monotonicity in monotone-TDLNM, due to the smoothing weight, makes exact coverage of a flat exposure-response impossible (e.g., below 25C during periods of nonzero effect) resulting in non-coverage at lower exposure-concentrations despite a roughly flat exposure response curve. TDLNM also maintains nominal coverage of the true effects but the credible interval width is generally larger than monotone-TDLNM. GAM models have adaquate coverage by 95\% confidence intervals, but the largest interval width. The wiggliness of splines in GAM contributes to additional uncertainty across the exposure-lag-response function, especially near boundaries where spline-based methods have the largest uncertainty. Addition targeted penalization to GAM decreases average width substantially, making it similar to TDLNM. 

Our proposed method generally has the highest precision compared to TDLNM and GAM across all scenarios and error settings. Compared to GAM, TDLNM also has higher precision on average. The inclusion of targeted penalization in GAM increases precision slightly. With respect to true positives (probability of detecting a correct period of nonzero effect), GAM consistently outperforms the other methods with the trade off of the largest false positive rate (probability of misidentifying an effect). The addition of informative prior information improves the true positive rate for monotone-TDLNM with larger differences occurring at higher error settings. Changing the splitting probabilities for TDLNM does not have a large effect on the model outcomes.

The Gelman-Rubin statistic, denoted $\hat{R}$, for convergence was calculated at each point on the exposure-lag-response surface and the median was preserved. We present the average $\hat{R}$ across simulation replicates for each scenario and error combination in Supplementary Materials \citep{MorkBASupp}. We note that $\hat{R}$ is near 1 for both monotone-TDLNM and TDLNM, indicating good convergence.

\section{Temperature related mortality}
We illustrate our method by analyzing the association between extreme heat or cold and daily mortality. Let $d_t$ represent the number of deaths on day $t$ while temperature during the past 20 days is given by $x_t,x_{t-1},\ldots,x_{t-20}$. We use mortality and temperature data from the National Morbidity, Mortality and Air Pollution Study, in the city of Chicago between 1987 to 2000 \citep{Samet2000TheIssues.}. The data is publicly available in the \texttt{R} package \texttt{dlnm}. Because extreme hot and cold temperatures may each increase mortality and break the monotonicity assumption in our model, we separately analyze summer (May-September) and winter (October-April) time periods. Days were removed from the analysis if lagged temperature was missing, resulting in 2,142 summer days and 2,952 winter days. Separately for summer and winter, we fit the model
\begin{equation}
    \mathbb{E}[\log(d_t)]=\gamma_{my}(t)+\delta_{dow}(t) +\sum_{l=0}^{20} w(x_{t-l},l)
\end{equation}
where $\gamma_{my}(t)$ is an intercept for month and year at time $t$, $\delta_{dow}(t)$ is an intercept for the day of the week. We assume a constant variance for the log-rate death outcome conditional on the time and temperature, i.e., $\text{Var}[\log(d_t)|t,\mathbf{x}_t]=\sigma^2$. We reversed the sign of temperature in the winter model so that $w(x_{t-l},l)$ is monotone increasing for monotone-TDLNM. Assumptions of constant variance and identically distributed normal errors are considered in Supplementary Materials Section 2 \citep{MorkBASupp}. Good convergence was demonstrated by the tree-based models, as indicated by traceplots and the Gelman-Rubin statistic; details are available in Supplementary Materials Section 2.

We applied informative priors reflecting the strong body of evidence that heat up to five days prior is related to mortality \citep{baccini2008heat,yu2012daily,ragettli2017exploring}. To incorporate informative prior information in the summer monotone-TDLNM approach, we applied the informative prior described in our simulation study to the first six lag periods ($l=0,\ldots,5$). For the winter temperature and mortality analysis we used a vague prior to reflect the added uncertainty around cold-temperature related mortality. Specifically, in the winter analysis we set $\gamma_{0l}$ and $\Sigma$ such that 95\% of the time $\pi_0(\gamma_{0l})$ lies between 0.005 and 0.995. As a sensitivity we repeated the summer data analysis with the same vague prior.

We compare results to TDLNM (no monotonicity assumption) and penalized spline DLNM (GAM) \citep{Mork2022TreedModels, Gasparrini2017}. The treed models were run for 5,000 MCMC iterations thinned by 10 after 2,000 burn-in iterations. We combined posterior samples from 5 independent Markov chains. For GAM, we specify third degree B-splines with 10 degrees of freedom in both lag and temperature dimensions and second order difference penalties. In the GAM summer model we added varying ridge penalization to the last seven degrees of freedom in the lag dimension as an informative prior (see e.g., \citet{Gasparrini2017} for more details). In both summer and winter mortality data, we estimate temperature-mortality relationships from the DLNM relative to 20 degrees Celsius and back transform results to estimated percent change in mortality at a given lag-temperature combination.

\begin{figure}[!ht]
    \centering
    \includegraphics[width=10cm]{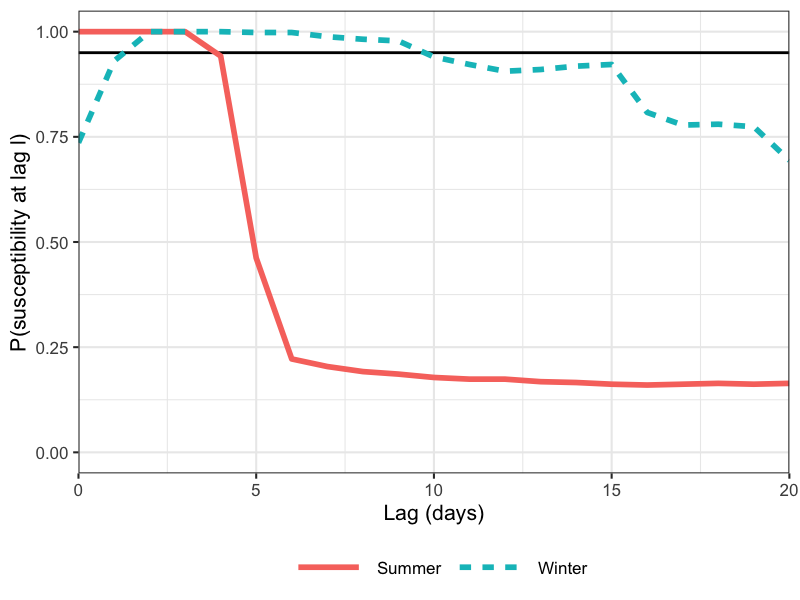}
    \caption{Posterior probability of susceptibility at a given lag, for summer (solid line) and winter (dashed line) models.}
    \label{fig:prob_effect}
\end{figure}

Monotone-TDLNM allows us to compute a posterior probability of non-zero temperature related mortality during a given lag period. The posterior probabilities of effect are shown in Figure \ref{fig:prob_effect}. We use posterior probability above $0.95$ to indicate a nonzero effect. The results from monotone-TDLNM indicate a relationship for heat-related (summer) mortality during lags $0-3$ days prior and cold-related (winter) mortality during lags $1-9$ days prior. Due to the monotonicity assumption in monotone-TDLNM we can specifically say these relationships are due to heat or cold. TDLNM and GAM do not share a similar variable selection method and we identify possible periods of susceptibility by identifying where 95\% credible intervals do not contain zero. Because the exposure-response functions in TDLNM and GAM are not monotone, the identified periods of susceptibility may be due to increased or decreased temperatures and requires inspection of the posterior distribution of the exposure-time-response surface. Based on the credible intervals, TDLNM indicates nonzero relationships between mortality and temperature in summer at lags $0-11$ and 19 days prior and for winter temperatures at lags $0-10$ days prior; GAM finds nonzero relationships between mortality and summer temperatures during lags $0-6$, $8-11$ and 13 days prior, and due to winter temperatures $0-5$ and $7-9$ days prior.

\begin{figure}[!ht]
     \centering
     \begin{subfigure}[b]{10cm}
         \centering
         \includegraphics[width=\textwidth]{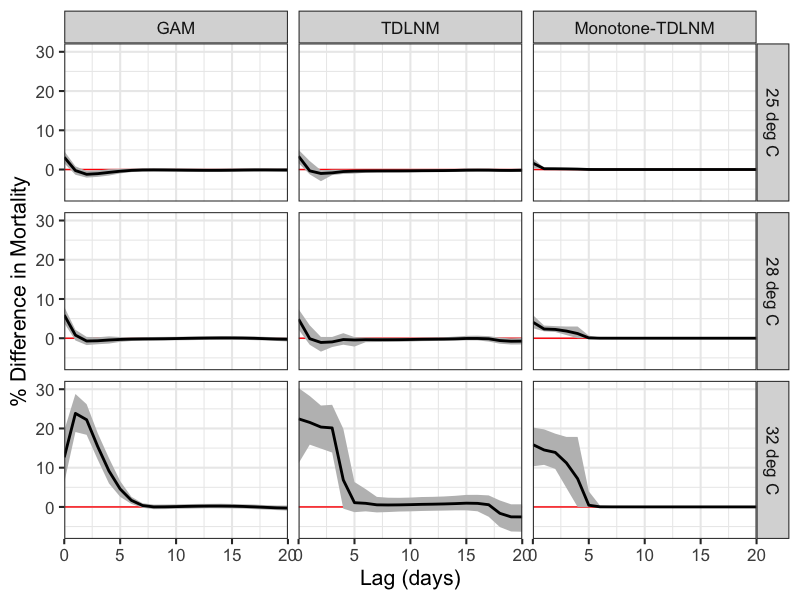}
         \caption{Lagged effects of heat-related mortality at three different temperature levels (rows) as estimated by three DLNM methods (columns). The x-axis indicates the lag time in days while the y-axis is unique to each row and describes the estimated percent change in mortality. The dark black lines show the estimated mean relationship while the gray area describes the 95\% credible interval.}
     \label{fig:summer-temp}
     \end{subfigure}
     \begin{subfigure}[b]{10cm}
         \centering
         \includegraphics[width=\textwidth]{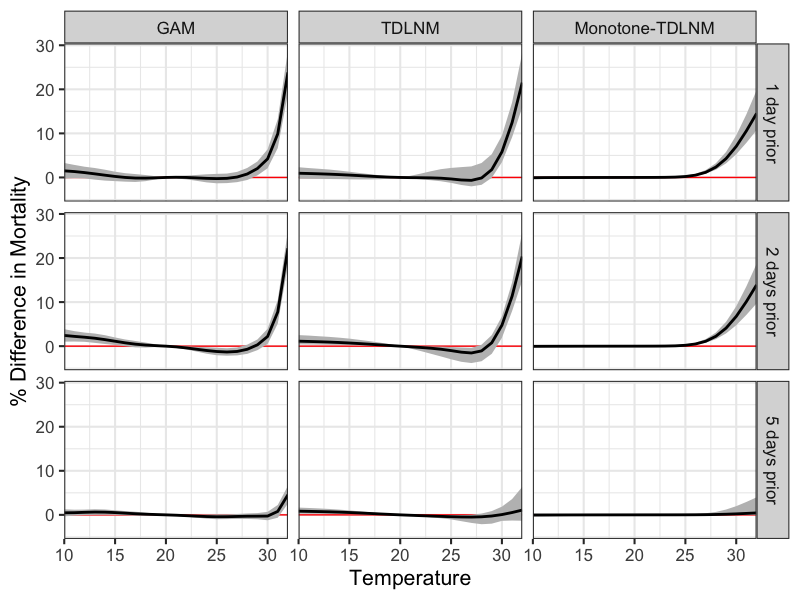}
         \caption{Temperature-specific effects of heat-related mortality at three different lag times (rows) as estimated by three DLNM methods (columns). The x-axis indicates the temperature in degrees Celsius while the y-axis is unique to each row and describes the estimated percent change in mortality. The dark black lines show the estimated mean relationship while the gray area describes the 95\% credible interval.}
     \label{fig:summer-lag}
     \end{subfigure}
     \caption{}
     \label{fig:summer}
\end{figure}

Figure \ref{fig:summer-temp} shows slices of the estimated DLNM at three different temperatures (25, 28, and 32 degrees C). This is the estimated percent difference in mean mortality for each temperature value compared to 20 degrees C by lag. Figure \ref{fig:summer-lag} shows slices at three different lag periods (1, 2, and 5 days prior) for each of the three compared methods. This is the exposure-response function between temperature and mortality on 1, 2, and 5 days post exposure. Only monotone-TDLNM shows longer lagged effects at lower temperatures (28 degrees C from $0-3$ days prior), the other methods show an immediate effect (same day of exposure) as well as intermittent nonzero relationships at longer lags (e.g., GAM shows increased mortality for temperatures below 20C during lags $8-11$ as well as decreased mortality above 20C at lag 13). For higher temperatures (32 degrees C), GAM shows effects extending back 6 days while the lagged relationships for TDLNM and monotone-TDLNM are similar but only extend to 3 days prior. When looking at slices in the lag dimension, all models indicate an exponential-like relationship between temperature and mortality. At 2 days prior, monotone-TDLNM indicates the relationship to increased mortality extends to lower temperatures than the other models. We also note that the credible interval widths for monotone-TDLNM are similar or smaller than the competing methods, especially at low temperatures and later lags where the effects are shrunk towards zero. The summer sensitivity analysis had similar findings, details are given in Supplementary Material Section 2 \citep{MorkBASupp}.

\begin{figure}[!ht]
     \centering
     \begin{subfigure}[b]{10cm}
         \centering
         \includegraphics[width=\textwidth]{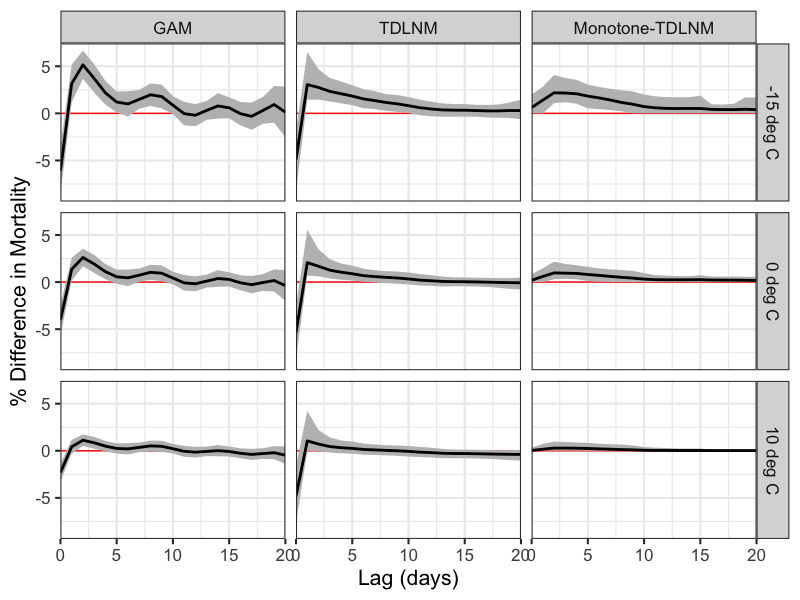}
         \caption{Lagged effects of cold-related mortality at three different temperature levels (rows) as estimated by three DLNM methods (columns). The x-axis indicates the lag time in days while the y-axis is unique to each row and describes the estimated percent change in mortality. The dark black lines show the estimated mean relationship while the gray area describes the 95\% credible interval.}
     \label{fig:winter-temp}
     \end{subfigure}
     \begin{subfigure}[b]{10cm}
         \centering
         \includegraphics[width=\textwidth]{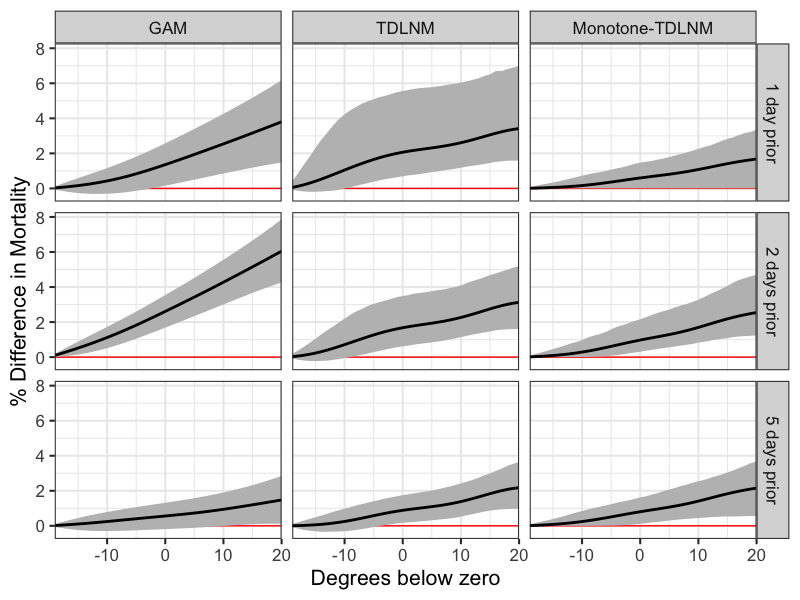}
         \caption{Temperature-specific effects of cold-related mortality at three different lag times (rows) as estimated by three DLNM methods (columns). The x-axis indicates the temperature in degrees Celsius below zero while the y-axis is unique to each row and describes the estimated percent change in mortality. The dark black lines show the estimated mean relationship while the gray area describes the 95\% credible interval.}
     \label{fig:winter-lag}
     \end{subfigure}
     \caption{}
     \label{fig:winter}
\end{figure}

Figure \ref{fig:winter-temp} shows the lagged relationship to mortality at temperatures -15, 0, and 10 degrees C. TDLNM and GAM indicate decreased mortality at lag 0 followed by increased mortality related to low temperatures during previous days. Monotone-TDLNM and TDLNM show potentially longer lagged relationships between lower temperatures and mortality compared to GAM. The wiggliness of GAM also indicates no relationship between mortality at cold temperatures at 6 days prior which is likely a spline-induced artifact. Figure \ref{fig:winter-lag} shows the exposure-response relationship between mortality and temperature (degrees below zero) at 1, 2, and 5 days prior. We see a more linear relationship between lower temperatures and mortality in all models where GAM extends up to temperatures near 20C while the tree-based methods only indicate increased mortality at lower temperatures.

\section{Discussion}

DLNMs have become a standard tool in environmental epidemiology. Most commonly researchers use spline-based DLNMs to estimate the association between an exposure and a lagged health outcome, such as temperature and mortality on the same day and following 20 days as we consider in this paper. In such analyses, it is rare to consider prior information on the shape of the exposure-lag-response function. This is despite ample prior research that shows that the exposure-response function is monotone for many exposure-response pairs or sheds light on which lags are likely to be associated with the outcome. The decision to not include information on monotonicity is in large part due to a lack of available statistical methods to estimate monotone DLNMs.

We propose a regression tree based DLNM that is constrained to have a monotone exposure-response relationship at each lag time. Our work builds of the popular spline-based DLNM \citep{Gasparrini2017} and existing tree-based TDLNM \citep{Mork2022TreedModels}, both of which impose no constraints on the shape of the exposure-response relationship. We propose a nested-tree approach that uses one tree to partition of lag period into discrete time segments and a set of nested trees that capture the exposure-response relationship during each time segment. The approach can be thought of as a Bayesian tree analog to the crossbasis DLNM that uses two basis expansion, one in the lag direction and a second in the exposure-concentrations direction, to estimate a DLNM \citep{Gasparrini2010}. 

Our proposed monotone-TDLNM allows for easy inclusion of prior information and inference on lags for which there is a nonzero exposure-response relationship through Bayesian variable selection methods incorporated into a regression tree setting. Identifying lags with nonzero exposure-response relationships is challenging with previous implementations of DLNM. Similar time selection methods have been proposed for linear DLMs \citep{Warren2020CriticalBirth}, but no such methods are available for nonlinear DLNMs. No previous methods have considered inclusion of prior information on which lags are associated with an outcome.

Through simulation, we show that the monotone-TDLNM resulted in more precise estimation of the exposure-lag-response function and lag selection compared to a penalized spline DLNM and unconstrained TDLNM. The monotone-TDLNM resulted in smaller credible intervals that still maintained the nominal coverage level. In an analysis of summer heat exposure and winter cold exposure and mortality in a Chicago, Illinois, USA time-series study we found that both summer heat and winter cold were associated with increased mortality. Importantly, the unconstrained models were consistent with a monotone relationship and such a relationship has been found in other analyses of temperature and mortality. However, the constrained model that includes prior information on monotonicity resulted in more precise estimates of the exposure-lag-response function. This highlights one of the advantages of the proposed model. In addition, the easy inference on the lags with nonzero effect highlights a second major advantage over previous methods. 

A clear limitation of the proposed approach, or any constrained regression method, is that violation of the monotonicity assumption would be a major misapplication and result in bias and incorrect inference. In the case of temperature and mortality this monotonicity by season is well established. However, in other situations it is important to check these assumptions. For example, a researcher may consider also fitting unconstrained models and looking for evidence of departures from monotonicity as we demonstrated in our data application. 

Our work adds to a recent literature on incorporating prior information in environmental epidemiology analyses. \cite{Thomas2007} provides a compelling argument for incorporating biological prior information in Bayesian analyses of the health effects of environmental exposures. Recent work has promoted the use of informative priors in several model types \citep{reich2020integrative, McGee2022}, including previous models that impose monotonicity in the exposure-response relationship \citep{Powell2012,Wilson2014c}. As was so well stated by \citep{Thomas2007}, ``by directly incorporating into our analyses information from other studies or allied fields---we can improve our ability to distinguish true causes of disease from noise and bias.''

\clearpage
\bibliographystyle{ba}
\bibliography{reference_combined}

\begin{acks}[Acknowledgments]
Research reported in this publication was supported by National Institute of Environmental Health Sciences of the National Institutes of Health under award number R01ES029943 and National Institute of Aging of the National Institutes of Health under award number R01AG066793.
\end{acks}

\begin{supplement}
\stitle{Supplementary Materials to: Incorporating prior information into distributed lag nonlinear models with zero-inflated monotone regression trees}
\sdescription{This document includes additional simulation and data analysis results and figures.}
\end{supplement}

\end{document}